\documentclass[12pt,preprint,aps,nofootinbib,shownopacs]{revtex4}
\usepackage{graphicx}
\usepackage{psfig}
\newcommand*{\be}{\begin{equation}}
\newcommand*{\ee}{\end{equation}}
\newcommand*{\bea}{\begin{eqnarray}}
\newcommand*{\eea}{\end{eqnarray}}
\providecommand*{\ler}{\stackrel{\scriptstyle <}{\scriptstyle \sim}}
\providecommand*{\ger}{\stackrel{\scriptstyle >}{\scriptstyle \sim}}
\newcommand{\nn}{\nonumber}
\newcommand{\Frac}[2]{\frac{\displaystyle{#1}}{\displaystyle{#2}}}
\newcommand{\lsim}{\raise0.3ex\hbox{$\;<$\kern-0.75em\raise-1.1ex\hbox{$\sim\;$}}}
\newcommand{\gsim}{\raise0.3ex\hbox{$\;>$\kern-0.75em\raise-1.1ex\hbox{$\sim\;$}}}

\begin{document}
\preprint{OUTP-02-33P}
\title{Seesaw and Lepton Flavour Violation in SUSY $SO(10)$}
\author{Antonio Masiero}
\email{masiero@pd.infn.it}
\affiliation{Dip. di Fisica `G. Galilei', Univ. di Padova  and  
 INFN, Sezione di Padova, via F. Marzolo 8,  I-35131, Padua, Italy.} 
\author{Sudhir K. Vempati}
\email{vempati@pd.infn.it}
\affiliation{Dip. di Fisica `G. Galilei', Univ. di Padova  and  
 INFN, Sezione di Padova, via F. Marzolo 8,  I-35131, Padua, Italy.} 
\author{Oscar Vives}
\email{vives@thphys.ox.ac.uk}
\affiliation{Dep. of Physics, U. of Oxford, 1 Keble Road, Oxford, OX1 3NP, UK.}

\begin{abstract}
That $\mu \rightarrow e, \gamma$ and $\tau \rightarrow \mu,\gamma$ are
sensitive probes of SUSY models with a see-saw mechanism is a well
accepted fact. Here we propose a `top-down' approach in a general SUSY
$SO(10)$ scheme. In this framework, we show that at least one of the
neutrino Yukawa couplings is as large as the top Yukawa coupling.
This leads to a strong enhancement of these leptonic flavour changing
decay rates.  We examine two `extreme' cases, where the lepton mixing
angles in the neutrino Yukawa couplings are either small (CKM-like) or
large (PMNS-like).  In these two cases, we quantify the sensitivity of
leptonic radiative decays to the SUSY mass spectrum. In the PMNS case,
we  find that the ongoing experiments at the B-factories can completely probe
the spectrum up to gaugino masses of 500 GeV (any tan $\beta$).  
Even in the case of CKM-like mixings, large regions of the parameter space 
will be probed in the near future, making these two processes leading 
candidates for indirect SUSY searches. 
\end{abstract}

\maketitle

\section{Introduction}

It is known that Flavour Changing Neutral Current (FCNC) processes
play an important role in the search for indirect signals of new
physics \cite{fcncreview}.  On the other end the accumulating
concordance between the Standard Model (SM) expectations and the vast
range of experimental results in FCNC and $CP$ violation point towards
a low energy new physics which is flavour blind.  For instance
considering Supersymmetric (SUSY) extensions of the SM all this
phenomenological evidence could be seen as a consequence of the fact
that the mechanism that breaks SUSY and conveys SUSY breaking to the
observable sector is completely flavour blind. If this is indeed the
case, it might be difficult to see any deviation at all from the SM
predictions through FCNC \cite{flavourblind}. However, we already know
that the SM is not enough in that it has to be supplemented by a
mechanism to provide non-zero neutrino masses and mixings.  An
appealing example of this is the seesaw mechanism \cite{seesaw} where
we introduce neutrino Yukawa couplings $h^\nu$, and heavy right-handed
neutrino Majorana masses $M_R$.

In spite of the possible flavour-blindness of SUSY breaking, the
Supersymmetrization of the Seesaw leads to new SUSY sources of Lepton
Flavour Violation (LFV) \cite{amfb}. The point is that in the running of the
slepton mass matrices down to the electroweak scale, contributions to these
matrices proportional to $h^\nu h^{\nu\,\dagger}$ give rise to a mismatch
in the diagonalisation of the lepton-slepton mass matrices and hence
the appearance of flavour changing gaugino-lepton-slepton vertices.

This effect was already noticed several years ago. What is new at this moment
is the improvement in our knowledge of neutrino physics and prospects of
better bounds from LFV physics. 
However, as shown in several recent studies 
\cite{casasibarra,lavignac1,lavignac2},
the low energy data on neutrino masses and mixings by themselves cannot
predict the decay rates of $l_j \to l_i \gamma$ for a given set of 
SUSY breaking parameters. This ambiguity can be traced to the fact that the 
unknown parameters of $h^\nu$ and $M_R$ cannot be completely fixed even after
knowing all the three masses and three mixing angles of the neutrino 
sector. 
Thus any ``bottom--up'' approach to the study of these lepton flavour 
violating processes suffers a large ambiguity which is encoded in an 
arbitrary orthogonal matrix $R$ \cite{casasibarra,lavignac1,lavignac2} relating 
these low energy neutrino parameters to the unknown Seesaw parameters $h^\nu$
and $M_R$\footnote{It is possible that mechanisms other than the seesaw can 
be more predictive for Lepton Flavour Violation. 
See, for an example, Ref.\cite{arossi}.}. 
Several interesting works \cite{others} have studied this 
problem in detail for various neutrino spectra - hierarchical, 
inverse-hierarchical, degenerate - and under various assumptions on the mass 
eigenvalues of the right handed neutrino mass matrix. 
The bottom line of these analysis is that $l_j \to l_i \gamma$ and neutrino 
oscillation experiments provide complementary pieces of information to 
determine the seesaw parameter space \cite{ibarradavidson}. 

In this paper, we take an alternative point of view namely, we use a
`top-down' approach introducing a high energy framework which
eliminates the ambiguity on the Seesaw parameters. We consider a SUSY
$SO(10)$ model where all fermions in a generation are included in a
single representation. The crucial find of our analysis is that in a
\textit{generic} $SO(10)$ model at least one of the neutrino Yukawa
couplings is of the same order as the large top Yukawa coupling. 
By \textit{generic} we mean here that the result holds irrespective 
of the representations of
the Higgs fields used to generate masses for the fermions. However, as
expected the complete structure of $h^\nu$ remains unpredictable in such 
a general case. Essentially, the specification of the Grand Unified gauge 
group is not sufficient to uniquely determine the mixings appearing in the 
diagonalisation of $h^\nu h^{\nu\,\dagger}$ relevant for the analysis of
$l_j \rightarrow l_i, \gamma$ decays. 

Motivated by simple $SO(10)$ models, we propose that the mixings diagonalising
the combination $h^\nu h^{\nu \dagger}$ would have magnitudes within the two 
`extreme' limits - namely, the CKM angles of quark mixing and the PMNS angles of
the leptonic mixing. We show semi-complete $SO(10)$ models where these two
`extreme' cases can be realised. We then demonstrate that phenomenologically 
viable seesaw mechanisms can be realised in both these cases by a suitable
choice of the right handed neutrino Majorana mass matrix. We believe that these
extreme cases would serve as ``benchmark" scenarios for the seesaw induced 
lepton flavour violation within the context of SUSY $SO(10)$.

For both these ``benchmark" scenarios, we have computed 
the $l_j \to l_i \gamma$ decay rates in a 
minimal supergravity (mSUGRA) or constrained MSSM (CMSSM) scenario, 
\textit{i.e,} assuming completely universal soft SUSY breaking. 
We find that the present and future observational
limits on $\mu \to e \gamma$ can significantly constrain large regions
in the parameter space even in the above mentioned small mixing (CKM) case.
In the more optimistic case of large mixing (PMNS) the decay rates can become
very high. Indeed, if the new proposals to explore LFV, mainly the $\mu \to 
e \gamma$ decay at PSI \cite{psi}, reach
the planned sensitivity these LFV processes are going to be fully 
complementary with SUSY searches at LHC.

The rest of the paper is organised as follows. In the next section we
show that in $SO(10)$ models at least one of the neutrino Yukawa
couplings is forced to be as large as the top Yukawa. Semi-complete 
$SO(10)$ models are presented in section III for the minimal and maximal
mixing scenarios. Results from numerical analysis and the constraints 
on SUSY parameter space are presented in section IV. Summarising remarks
are presented in the last section. 

\section{$SO(10)$ and Neutrino Yukawa Couplings} 


As mentioned in the introduction, the top-down approach makes it
possible to predict the neutrino Yukawa couplings removing the
ambiguity associated with the seesaw parameters relevant for lepton
flavour violation. In first place, we must choose a gauge group for
the Grand Unified Theory (GUT).  $SO(10)$ is the natural choice as it
is the minimal group that includes right handed neutrinos besides the
rest of the SM fermions in a single representation.  One would then
have to consider a particular choice of the Higgs fields
representations to generate fermion masses, which again introduces
some sort of an `ambiguity' at the high scale. To keep the discussion
as general as possible, we do not resort to any particular model of
fermion masses within $SO(10)$, but try to see how well one can
predict $h^\nu$ in a generic scenario.

In the $SO(10)$ gauge theory, all the known fermions and the 
right handed neutrinos are unified in a single representation 
of the gauge group, the \textbf{16}. To analyse the Yukawa matrices 
in this framework,  we need to specify the
superpotential.  In principle, the superpotential can receive contributions
both from renormalisable and non-renormalisable terms \cite{bhs}. The 
product of two \textbf{16} matter representations can only couple to 
\textbf{10}, \textbf{120} or \textbf{126} representations which can be 
formed either by a single Higgs field representation or a 
non-renormalisable product of representations of several Higgs fields.  
In either case, the  Yukawa matrices resulting from the couplings to 
\textbf{10} and \textbf{126} are complex symmetric whereas they are 
anti-symmetric when the couplings are to the \textbf{120}. 

Therefore, the most general $SO(10)$ superpotential relevant for
fermion masses can be written as
\be
W_{SO(10)} = h^{10}_{ij} 16_i~ 16_j~ 10 + h^{126}_{ij} 16_i~ 16_j~ 126 
+ h^{120}_{ij} 16_i~ 16_j~ 120,  
\ee
where $i,j$ refer to the generation indices. In terms of the SM fields, 
the Yukawa couplings relevant for fermion masses are given by \cite{strocchi}:
\bea
16\ 16\ 10\ &\supset& 5\ ( u u^c + \nu \nu^c) + \bar 5\
(d d^c + e e^c), \nn\\
16\ 16\ 126\ &\supset& 1\ \nu^c \nu^c + 15\ \nu \nu +
5\ ( u u^c -3~ \nu \nu^c) + \bar{45}\ (d d^c -3~ e e^c), \nn\\
16\ 16\ 120\ &\supset& 5\ \nu \nu^c + 45\ u u^c +
\bar 5\ ( d d^c + e e^c) + \bar{45}\ (d d^c -3~ e e^c),
\label{su5content}
\eea
where we have specified the corresponding $SU(5)$  Higgs representations 
for each  of the couplings and all the fermions are left handed fields. 
From above, it is clear that if only the \textbf{10} and 
\textbf{126} Higgs
representations are present in the theory, the Yukawa matrices of the
down quarks and charged leptons as well as the up quarks and neutrinos
are deeply related. In fact, in a model where only the {\bf 10}-plets 
are present we would have exact quark--lepton Yukawa unification, not
only among charged leptons and down quarks ($b$--$\tau$ unification) but 
also for up quarks and (Dirac) neutrinos. Similarly  a dominant contribution
from the {\bf 126} representation would predict as well quark--lepton 
unification, although introducing the Georgi--Jarlskog factors of 3
\cite{georgijarlskog}. 
If only one of these  representations or any combination of them contributes, 
the Yukawa matrices would be exactly symmetric. 
These properties are broken by the introduction of the {\bf 120}
representation. The Yukawa couplings of this representation are
anti-symmetric and break the quark--lepton unification because they can
contribute independently to the quark and lepton Yukawa matrices. 
In general, both the symmetric and anti-symmetric contributions can be present
leading to Yukawa matrices of generic nature. 

The resulting mass matrices can be written as
\bea
\label{upmats}
M^u &= & M^5_{10} + M^5_{126} + M^{45}_{120}, \\
\label{numats}
M^\nu_{LR} &= & M^5_{10} - 3~ M^5_{126} + M^{5}_{120}, \\
\label{downmats}
M^d &= & M^{\bar{5}}_{10} +  M^{\bar{45}}_{126} + M^{\bar{5}}_{120} 
+ M^{\bar{45}}_{120}, \\
\label{clepmats}
M^e &= & M^{\bar{5}}_{10} - 3  M^{\bar{45}}_{126} + M^{\bar{5}}_{120} 
- 3 M^{\bar{45}}_{120}, \\
M^\nu_{LL} &=& M^{15}_{126}, \\
M^\nu_{R} &=& M^{1}_{126}.
\eea

We come now to discuss the main result of this section: \textit{At least one
of the Yukawa couplings in $h^\nu~ =~ v_u^{-1}~M^\nu_{LR}$ has 
to be as large as the top Yukawa coupling}.  This result holds 
true in general independently from
the choice of the Higgses responsible for the masses in
Eqs.~(\ref{upmats}, \ref{numats}) provided that no accidental fine
tuned cancellations of the different contributions in Eq.~(\ref{numats}) are
present. If contributions from the \textbf{10}'s solely 
dominate, $h^\nu$ and $h^u$ would be equal. If this occurs for the 
\textbf{126}'s, then $h^\nu =- 3~ h^u$. 
In case both of them have dominant
entries, barring a rather precisely fine tuned
cancellation between $M^5_{10}$ and $M^5_{126}$ in
Eq.~(\ref{numats}), we expect at least one large entry to be present in $h^\nu$.
A dominant antisymmetric contribution to top quark mass 
due to the {\bf 120} Higgs is phenomenologically excluded since it would  
lead to at least a pair of heavy degenerate up quarks. However, the 
{\bf 120} can still provide a non negligible contribution to the up quark 
masses and in particular, to that of the top quark, if there is at the same
time a large symmetric contribution. In this case the up quark and
neutrino Yukawa matrices have both large symmetric and antisymmetric
contributions. The above stated result holds also in this general situation
(for a complete proof, see the Appendix).

Apart from sharing the property that at least one eigenvalue of both $M^u$ 
and $M^\nu_{LR}$ has to be large, for the rest it is clear from 
(\ref{upmats}) and  (\ref{numats}) that these two matrices are not aligned 
in general, and hence we may expect different mixing angles appearing 
from their diagonalisation. This freedom is removed if one sticks to 
particularly simple choices of the Higgses responsible for up quark and
neutrino masses. For instance, as long as one sticks to representations of
the Higgs fields which preserve the quark-lepton Yukawa unification, like
the \textbf{10}-plets (due to the underlying Pati-Salam symmetry), the mixing 
angles appearing in the diagonalisation of the up-quark mass matrix
 also appear in $h^\nu$. In this case $h^\nu$ can be 
completely predicted at the high scale. However, in a general scenario, this
need not hold true. Large contributions from \textbf{120} can break this
alignment between up and the neutrino Yukawa matrices as is evident from
Eqs.~(\ref{upmats} - \ref{clepmats}) and hence $h^\nu$ would be no longer 
predictable in this general situation. 
  
Keeping with our philosophy to be as general as possible, we find two
cases which would serve as `benchmark' scenarios for seesaw induced
lepton flavour violation in SUSY $SO(10)$. The first one corresponds
to a case where the mixing present in $h^\nu$ is small and
CKM-like. This is typical of the models where fermions attain their
masses through $10$-plets. We will call this case, `the minimal
case'. As a second case, we consider scenarios where the mixing in
$h^\nu$ is no longer small, but large like the observed PMNS
mixing. We will call this case the `the maximal case'.  Within
$SO(10)$ this is possible in models with asymmetric Yukawa
matrices. These two `benchmark' cases, we believe, would span the
range of lepton flavour violation generated by $h^\nu h^{\nu \dagger}$
at the weak scale in a fairly general way. We now proceed to study
these constraints for the two cases.

\section{$SO(10)$ and Lepton Flavour Violation} 

As mentioned in the Introduction, the off-diagonal entries in the slepton
mass matrices are generated by the product $h^\nu h^{\nu~T}$ through 
renormalisation group evolution\footnote{We have implicitly assumed mass 
matrices to be real for the following discussion and thus neglect any 
effects due to the presence of phases in the mass matrices.}. Having known 
the matrix $h^\nu$ or even
the product $h^\nu h^{\nu~T}$ at the high scale, the off-diagonal entry 
$[m_{\tilde{L}}^2]_{ij}$ can be calculated at the weak scale. These entries 
would then contribute to the lepton flavour violating decays 
$l_j \rightarrow l_i, \gamma$. A naive estimate of the branching ratios
of these decays is given by (in standard notation) \cite{hisano}: 
\be
\mbox{BR} (l_j \to l_i \gamma) \approx \Frac{ \alpha^3 ~
(~[m_{\tilde{\scriptscriptstyle{L}}}^2]_{ij} )^2 }
{G_F^2~ m^8_{SUSY}} \tan^2 \beta, 
\label{BR}
\ee 
where $m_{SUSY}$ represents the typical soft Supersymmetric
breaking mass. The present experimental limits on the
branching ratios of these decays are given as \cite{mega,belle}
\bea
\label{megprelimit}
BR(\mu \rightarrow e, \gamma)&\leq& 1.2 \times 10^{-11},\\
\label{tmgprelimit}
BR(\tau \rightarrow \mu, \gamma)&\leq& 5. \times 10^{-7}.
\eea 
In future these bounds are expected to improve at least by a few orders of 
magnitude. In particular, in the proposed experiment at PSI, the
limits on the $BR(\mu \rightarrow e, \gamma)$ are expected to improve to 
\cite{psi}
\be
BR(\mu \rightarrow e, \gamma) \leq 10^{-14}.  
\ee 
These limits would now constrain the parameters governing the sparticle
spectrum, namely $m_0, M_{1/2}, A_0, \mbox{sg}(\mu)$ and tan$\beta$ in the
mSUGRA with electroweak radiative breaking. It should be noted that the 
above mentioned RG effects would require 
the scale of Supersymmetry breaking to be higher than the scale of right 
handed neutrinos.
Thus in models of Gauge Mediated Supersymmetry Breaking (GMSB) these effects
would be absent. We now analyse these branching ratios in detail for the
two cases mentioned at the end of the previous section, namely, the small 
mixing case with CKM angles and the large mixing case with PMNS angles in $h^\nu$.  
\subsection{The minimal Case: \textit{CKM mixings in $h^\nu$}}
\label{sec:CKM} 
The minimal Higgs spectrum to obtain phenomenologically viable mass matrices
includes two \textbf{10}-plets, one coupling to
the up-sector and the other to the down-sector. In this way it is possible to 
obtain the required CKM mixing \cite{buchwyler} in the quark sector. 
The $SO(10)$
superpotential is now given by

\be
\label{primedbasis}
W_{SO(10)} = {1 \over 2}~  h^{u,\nu}_{ij} 16_i~ 16_j~ 10_u + 
{1 \over 2}~ h^{d,e}_{ij} 16_i ~16_j~ 10_d + 
{1 \over 2}~ h^R_{ij}~ 16_i~ 16_j~ 126. 
\ee
We further assume the {\bf 126} dimensional Higgs field gives
Majorana mass  \textit{only} to the right handed neutrinos. 
An additional feature of the above mass matrices is that all of them 
are \textit{symmetric}. Without loss of generality we can rotate the 
{\bf 16}-plet into a basis where the charged leptons and the down-type 
quarks are diagonal. 
In terms of the SM fields, we can rewrite the above as 
\be
W = h^u_{ij} ~Q_i~ u_j^c~ H_u + h^d_{ii}~ Q_i~ d_i^c~ H_d 
+ h^e_{ii}~ L_i~ e_i^c~ H_d + {h^\nu}_{ij}~ L_i ~\nu_j^c~ H_u 
+ {1 \over 2} {M_{R}}_{ij}~ \nu^c_i ~\nu^c_j. 
\ee
Immediately we see that the following mass relations hold between the quark
and leptonic mass matrices at the GUT scale\footnote{Clearly this relation 
cannot hold for the first two generations of down quarks and charged leptons.
As usual, small corrections due to  non-renormalisable operators or
suppressed renormalisable operators \cite{georgijarlskog} can be invoked.}: 
\be
\label{massrelations}
h^u  = h^\nu \;\;\;;\;\;\; h^d  = h^e . 
\ee
In the above basis, the symmetric $h^u$ is diagonalised by: 
\be
\label{htop}
V_{CKM}~ h^u~ V_{CKM}^{T} = h^u_{diag}. 
\ee
Hence from (\ref{massrelations}): 
\be
\label{hnumg}
 h^\nu = V_{CKM}^T~ h^u_{diag}~ V_{CKM}. \ee 
According to Eq.~(\ref{BR}), $\mbox{BR}(\mu \to e \gamma)$ depends on:
\be
\label{hnusqckm}
[h^\nu h^\nu]_{21} \approx h_t^2 ~V_{td}~ V_{ts} +
{\mathcal O}(h_c^2).  
\ee 
In this expression, the CKM angles are small
but the presence of the large top Yukawa coupling compensates for such
suppression. The large couplings in $h^\nu \sim {\mathcal O}(h_t)$ induce 
significant off-diagonal entries in $m_{\tilde L}^2$ through the RG
evolution between $M_{GUT}$ and the scale of the right-handed Majorana neutrinos
\footnote{Typically one has different mass scales associated with different
right handed neutrino masses.}, $M_{R_i}$. The induced off-diagonal entry 
relevant for $\mu \rightarrow e, \gamma$ is of the order,
\be
\label{ckmoffdapp}
[m_{\tilde{\scriptscriptstyle{L}}}^2]_{21} 
\approx -{1 \over 8 \pi^2} (3 m_0^2 + A_0^2) 
~h_t^2~ V_{td}~ V_{ts} \log ({M_{GUT} \over M_R}) + {\mathcal O}(h_c^2).
\ee 
The required right handed neutrino Majorana  mass matrix consistent 
with both the observed low energy neutrino masses and mixings as well 
as with CKM like mixings in $h^\nu$ is determined easily from the 
seesaw formula defined at the scale of right handed neutrinos as 
\bea
\label{seesaw}
m_\nu &=& - h^{\nu~T}~M_R^{-1}~ h^{\nu}~v_u^2,  \\
 &=& - h^{\nu}~M_R^{-1}~ h^{\nu}~v_u^2. 
\eea
where we have used the symmetric nature of the $h^\nu$ in the second equation.
Inverting  Eq.~(\ref{seesaw}), one gets:
\bea
\label{MRstrcture1}
M_R &=&  - h^\nu~ m_\nu^{-1}~ h^\nu ~v_u^2, \nonumber \\
\label{maar2}
 &=& V_{CKM}~ h_{diag}^u~ V_{CKM}^T~ m_{\nu}^{-1}
V_{CKM}~ h_{diag}^u~ V_{CKM}^T, 
\eea   
where we have used Eq.~(\ref{hnumg}) for $h^\nu$. Furthermore, 
$m_\nu^{-1}$ can be written as 
$m_\nu^{-1} = U_{PMNS}~ diag[m_{\nu}^{-1}]~ U_{PMNS}^T $, 
whose entries are determined at the low scale 
from neutrino oscillation experiments. The structure of $M_{R}$ can now be
derived\footnote{ The neutrino masses and mixings here are defined
at $M_{GUT}$. Radiative corrections can significantly modify the neutrino
spectrum at the weak scale \cite{nurad}. This is more true for the degenerate
spectrum of neutrino masses \cite{degenrad} and for some specific forms 
of $h^\nu$ \cite{antusch}. For our present discussion, with hierarchical
neutrino masses and up-quark like neutrino Yukawa matrices, we expect these
effects not to play a very significant role.}  
for a given set of
neutrino masses and mixing angles. Neglecting the small CKM mixing in $h^\nu$
we have
\be 
\label{ckmmr}
M_R \approx v_u^2~ \left( \begin{array}{ccc}
h_u^2 [m^{-1}_{\nu}]_{11} & 
h_u h_c [m^{-1}_{\nu}]_{12} & 
h_u h_t [m^{-1}_{\nu}]_{13}  \\
h_u h_c [m^{-1}_{\nu}]_{12} & 
h_c^2 [m^{-1}_{\nu}]_{22} & 
h_c h_t [m^{-1}_{\nu}]_{23}  \\
h_u h_t [m^{-1}_{\nu}]_{13}  & 
h_c h_t [m^{-1}_{\nu}]_{23} &
h_t^2 [m^{-1}_{\nu}]_{33} \end{array} \right).
\ee 
It is clear from above that the hierarchy in the $M_R$ mass matrix goes
as the square of the hierarchy in the up-type quark mass matrix.
Furthermore, for a hierarchical neutrino mass spectrum we have 
$m_{\nu_3} \approx \sqrt{\Delta m^2_{Atm}},~~
m_{\nu_{2}} \approx \sqrt{\Delta m^2_{\odot}}$ and $m_{\nu_{1}} \ll 
\sqrt{\Delta m^2_{\odot}}$ and for a nearly bi-maximal
$U_{PMNS}$ : 
\bea
U_{PMNS} \approx \pmatrix{1/\sqrt{2} & 1/\sqrt{2} & 0 \cr
-1/2 & 1/2 &1/\sqrt{2}  \cr
1/2 & -1/2 &1/\sqrt{2}}~,~~~~
\eea
it straight-forward to check that all the right handed neutrino mass 
eigenvalues are controlled by the smallest left-handed neutrino mass. 
\be
\label{mrapproxegckm}
M_{R_3} \approx {m_t^2 \over 4~ m_{\nu_1}} \;\;;\;\;
M_{R_2} \approx {m_c^2 \over 4~ m_{\nu_1}} \;\;;\;\;
M_{R_1} \approx {m_u^2 \over 2~ m_{\nu_1}} \;\;.\;
\ee 
This implies that we can not choose an arbitrarily small neutrino mass if
we want the right-handed neutrino masses to be below $M_{GUT}$. In our 
numerical examples in section IV, we choose $m_{\nu_3} = 0.05 \mbox{ eV},
m_{\nu_2} = 0.0055 \mbox{ eV}, m_{\nu_1} = 0.001 \mbox{ eV}$.

At this point we have both $h^\nu$ and $M_R$ determined and we can now 
use the  experimental bounds on BR($\mu \rightarrow e, \gamma$) to
constrain the SUSY parameter space. The only unknowns are the SUSY
breaking soft terms $m_0$, $M_{1/2}$, $A_0$ and $\mbox{sg} (\mu)$, $\tan 
\beta$. For instance, from Eqs.~(\ref{BR},\ref{ckmoffdapp}) we have 
\be 
m_0^4 \geq  {3 ~\alpha^3 \over 8~\pi^2~G_F^2 }~ |\log{M_X \over M_R}|^2 ~
{h_t^4 ~V_{td}^2 ~V_{ts}^2 \over B}  \tan \beta^2, 
\ee
where $B$ represents the experimental limit on the branching ratio 
$BR(\mu \rightarrow e, \gamma)$, $A_0$ is assumed to vanish and $m_0$ is
identified with $m_{SUSY}$. Taking
the futuristic limit of $B \leq 10^{-14}$, we see that $m_0$ can be
probed up to $1$ TeV for large $\tan \beta$ $\sim 40$, bordering the limit
that will be probed at LHC \cite{lhc}. We quantify these results
with a numerical analysis in the next section. 

\subsection{The maximal case: \textit{PMNS mixing angles in $h^{\nu}$} }
\label{sec:PMNS}
The minimal $SO(10)$ model presented in the previous sub-section would
inevitably lead to small mixing in $h^\nu$. In fact, with two Higgs fields
in symmetric representations, giving masses to the up-sector and the 
down-sector separately, it would be difficult to avoid the small CKM like
mixing in $h^\nu$. To generate mixing angles larger than CKM angles, asymmetric
mass matrices have to be considered. In general, it is sufficient to introduce
asymmetric textures either in the up-sector or in the down-sector. In the 
present case, we assume that the down-sector couples to a combination of Higgs 
representations (symmetric and anti-symmetric) \footnote{The couplings of $\Phi$ 
in the superpotential can be either renormalisable or non-renormalisable. 
See \cite{chang} for a non-renormalisable example.}
$\Phi$, leading to an asymmetric mass matrix in the basis where the up-sector
is diagonal. We have : 
\be
\label{mnsso10}
W_{SO(10)} = {1 \over 2}~  h^{u,\nu}_{ii}~ 16_i ~16_j 10^u + 
{1 \over 2}~ h^{d,e}_{ij}~ 16_i ~16_j \Phi + 
{1 \over 2}~ h^R_{ij}~ 16_i~ 16_j 126~, 
\ee 
where the \textbf{126}, as before, generates only the right handed neutrino
mass matrix. To study the consequences
of these assumptions, we see that at the level of $SU(5)$, we have
\be
W_{SU(5)} = {1 \over 2}~ h^u_{ii}~ 10_i ~10_i ~5_u 
+ h^\nu_{ii} ~\bar{5}_i~ 1_i~ 5_u + 
h^d_{ij}~ 10_i ~\bar{5}_j~ \bar{5}_d + {1 \over 2}~M^R_{ij}~ 1_i 1_j,
\ee 
where we have decomposed the $16$ into $10 + \bar{5} + 1$ and $5_u$ and
$\bar{5}_d$ are components of $10_u$ and $\Phi$ respectively. To have large
mixing $\sim~ U_{PMNS}$  in $h^\nu$ we see that the asymmetric matrix $h^d$
should now be able to generate both the CKM mixing as well as PMNS mixing. 
This is possible if 
\be
V_{CKM}^T~ h^d~ U_{PMNS}^T = h^d_{diag}. 
\ee 
This would mean that the $10$ which contains the left handed down-quarks would
be rotated by the CKM matrix whereas the $\bar{5}$ which contains the left
handed charged leptons would  be rotated by the $U_{PMNS}$ matrix to go into
their respective mass bases \cite{chang}. Thus we have, in analogy with the
previous sub-section, the following relations hold true in the basis where 
charged leptons and down quarks are diagonal:
\bea
h^u &=& V_{CKM}~ h^u_{diag}~ V_{CKM}^T~ ,  \\
\label{hnumns}
h^\nu &=& U_{PMNS}~ h^u_{diag}. 
\eea
Using the seesaw formula of Eq.~({\ref{seesaw}) and  Eq.~(\ref{hnumns}) we have
\be
M_{R} = \mbox{Diag}\{ {m_u^2 \over m_{\nu_1}},~{m_c^2 \over m_{\nu_2}},
~{m_t^2 \over m_{\nu_3}} \}. 
\ee 
This would mean that this setup would require $M_R$ to be diagonal at the 
$SO(10)$ level in the basis of diagonal $h^{u,\nu}$, Eq.~(\ref{mnsso10}). 
We now turn our attention to  lepton flavour violation in the scenario. The
branching ratio, BR($\mu \rightarrow e, \gamma)$ would now be dependent on: 
\be
\label{hnusqmns}
[h^\nu h^{\nu~T}]_{21} = h_t^2~ U_{\mu 3}~ U_{e 3} + h_c^2~ U_{\mu 2}~ U_{e 2} +
\mathcal{O}(h_u^2), 
\ee 
where $U_{fi}$ are elements of the $U_{PMNS}$ matrix. 
It is immediately clear from the above that in contrast to the CKM case here
the dominant contribution to the off-diagonal entries depends on the 
unknown magnitude of the element $U_{e3}$. If $U_{e3}$ is very close to its
present limit $\sim~0.2$\cite{chooz}, the first term on the RHS of the Eq.~(\ref{hnusqmns})
would dominate. Moreover, this would lead to large contributions to the 
off-diagonal entries in the slepton masses with $U_{\mu 3}$ of 
${\mathcal O}(1)$. We have
\be
\label{mnsoffdapp}
[m_{\tilde{\scriptscriptstyle{L}}}^2]_{21} 
\approx -{1 \over 8 \pi^2} (3 m_0^2 + A_0^2) 
~h_t^2~ U_{\mu 3}~ U_{e3} \log ({M_{GUT} \over M_R}) + {\mathcal O}(h_c^2).
\ee 
The above contribution is large by a factor $(U_{\mu 3} U_{e3})/ (V_{td} V_{ts})
\sim 140 $ compared to the CKM case. From Eq.~(\ref{BR}) we see that it would mean
about a factor $10^4$ times larger than the CKM case in 
BR($\mu \rightarrow e, \gamma)$. In case $U_{e3}$ is very small, \textit{i.e,} 
either zero or $\ler~ h_c^2/h_t^2~U_{e2}~ \sim 4 \times 10^{-5}$, the 
second term $\propto ~h_c^2$
in Eq.~(\ref{hnusqmns}) would dominate. However the off-diagonal contribution
in slepton masses, now being proportional to charm Yukawa could be much smaller, 
in fact, even  smaller than the CKM contribution by a factor 
\be{h_c^2~ U_{\mu 2} ~U_{e 2} \over  h_t^2 ~V_{td} ~V_{ts}} 
\sim  7 \times 10^{-2}.
\ee

If $U_{e3}$ is close to it's present limit, the current bound on
BR($\mu \rightarrow e, \gamma$) would already be sufficient to
produce stringent limits on the SUSY mass spectrum. Indeed from:

\be 
m_0^4 \geq  {3 ~\alpha^3 \over 8~\pi^2~G_F^2 }~ |\log{M_X \over M_R}|^2 ~
{h_t^4 ~U_{\mu3}^2 ~U_{e3}^2 \over B}  \tan \beta^2 ~~(A_0 = 0), 
\ee

$B \leq 10^{-11}$ probes $m_0$ at the TeV level
even for small $\tan \beta$. We make these statements more
concrete in the next section with our results from numerical analysis. 

\section{$SO(10)$ and SUSY mass spectrum } 

As mentioned in the introduction we have chosen to work within a
mSUGRA framework with flavour blind universal soft breaking terms
at the GUT scale.  It is known \cite{bhs} that if the soft
breaking terms are hard at scales above $M_{GUT}$ and if the
perturbative RG approach can be safely used in such energy interval
above $M_{GUT}$, then the large value of $h_t$ can lead to relevant
radiative contributions to the mass of the stau, spoiling the slepton
mass universality at $M_{GUT}$. Here we are interested in the effects
produced by the large $h^\nu$ coupling in the running of the
$m_{\tilde{L}}^2$ down to the $M_R$ scales which are below
$M_{GUT}$. For simplicity, we assume the soft masses to be universal
at $M_{GUT}$, hence neglecting the possible above mentioned effects
which would further enhance the stringent bounds that we obtained
here.

In our numerical analysis, we have considered MSSMRN 
(MSSM + Right handed neutrinos) RGE from the scale 
$M_{GUT}$ down to the scale of the different RH neutrinos, which we integrate 
out in several steps until the scale of the lightest RH neutrino, $M_{R_1}$. 
>From $M_{R_1}$ to $M_Z$ the standard MSSM RGE are used \cite{antusch}. 
As a result of these renormalisation group effects,
the slepton mass matrices which were diagonal at $M_{GUT}$ to start 
with are now non-diagonal
at the weak scale. At this latter scale, we have numerically diagonalised these 
mass matrices and the corresponding eigenvalues and mixing matrices are found.
We used the complete calculations of 
Hisano \textit{et. al} \cite{hisano} to 
compute the branching ratios of both $\mu \rightarrow e, \gamma$ and 
$\tau \rightarrow, \mu, \gamma$.

We produce scatter plots of BR($\mu \rightarrow e, \gamma$) and 
BR($\tau \rightarrow \mu,\gamma$) vs. $M_{1/2}$ for small and large values
of tan $\beta$. We restrict the allowed SUSY parameter space by imposing
i) experimental constraints from direct sparticle searches,
in particular requiring $m_{\tilde{l}}$ and $m_{\chi^\pm}$ to be above $M_{Z}$
and $m_{\tilde{\nu}} \geq 33$ GeV; ii) the LSP to be neutral and
iii)~$2 \times 10^{-4} \leq $~BR($b \rightarrow s, \gamma$)$~ 
\leq 4 \times 10^{-4}$. In each scatter plot, for a given tan $\beta$ 
and $sg(\mu)$, the rest of the parameters are allowed to vary 
within the ranges: \\
a) 90 GeV  $\leq ~m_0~\leq$ 900 GeV ; \\ 
b) 90 GeV  $\leq ~M_{1/2}~\leq$ 700 GeV, \\
and  c) $A_0$ is parameterised as $ A_0~ =~ A_1~ m_0$ with $A_1$
lying between (-3,~3). For the neutrino mass eigenvalues, we have used
the values specified at the end of section IIIA. We have chosen 
$\sin^2 2 \theta_{\mbox{solar}} ~\approx 0.71;~ 
\sin^2 2 \theta_{\mbox{atm}} ~\approx 1$ and $U_{e3} ~\approx 0.15$. 

In Fig. 1a) and 1b) we show the scatter plots for BR($\mu \rightarrow
e,\gamma$ ) for the CKM case and $\tan \beta = 2$ and $\tan \beta =
40$ respectively. Similar figures for BR($\tau \rightarrow \mu,\gamma$
) and the same values of $\tan \beta$ are presented in Figs. 1c) and
1d). All these plots are calculated with $\mu > 0$ but the results do
not change significantly with negative $\mu$. These plots reflect 
an interesting correlation between the branching ratios
and the GUT value of the universal gaugino mass. This is due to the
fact that the gaugino mass fixes the chargino and neutralino masses at
$M_W$ and, to a small extent it also influences the slepton masses
through RGE. However, for a fixed $M_{1/2}$ the different values of
$m_0$ and $A_0$ can change the value of the BR within a range of 3
orders of magnitude. Nevertheless, this fact has still important
consequences.  For instance, for $\tan \beta=40$ reaching a
sensitivity of $10^{-14}$ for BR$(\mu \to e \gamma)$ would allow us to
probe completely the SUSY spectrum up to $M_{1/2} = 300$ GeV (notice
that this corresponds to gluino and squark masses of order 750 GeV)
and would still probe a large regions in parameter space up to
$M_{1/2} = 700$ GeV.  In the case of smaller values of $\tan \beta$
the BR scales as $(\tan \beta)^2$ and therefore for $\tan \beta = 2$
only a small part of the parameter space with $M_{1/2} \leq 300$ GeV
can be probed with this sensitivity. Similarly in the $\tau \to \mu \gamma$
decay a sensitivity of $6 \times 10^{-8}$ which could be reached in the B
factories in the near future \cite{belle,hisanonew}, would allow to probe a
sizeable piece of the parameter space for large $\tan \beta$.

In the PMNS scenario Figs. 2a) and 2b) show the plots for BR($\mu
\rightarrow e, \gamma$) for tan $\beta$ = 2, 40, whereas Figs. 2c and
2d are for BR($\tau \rightarrow \mu, \gamma$).  As we said in the
previous section (see Eq.~(\ref{mnsoffdapp})) in the PMNS case, the
results concerning BR($\mu \rightarrow e, \gamma$) strongly depend on
the unknown value of $U_{e3}$. In the plots 2a and 2b, the value of
$U_{e3}$ chosen is very close to the present experimental upper limit
\cite{chooz}. As long as $U_{e3} \ger 4 \times 10^{-5}$, the plots
scale as $U_{e3}^2$, while for $U_{e3} \ler 4 \times 10^{-5}$ the term
proportional to $m_c^2$ in Eq.~(\ref{mnsoffdapp}) starts dominating
and then, the result is insensitive to the choice of $U_{e3}$.  Here
we see that with the present limit on BR($\mu \rightarrow e, \gamma$),
all the parameter space would be completely excluded up to $M_{1/2}=
700$ GeV for $U_{e3} =0.15$ for any value of $\tan \beta$. For
instance, a value of $U_{e3} =0.01$ would reduce the BR by a factor of
225 and still most of the parameter space for $\tan \beta = 40$ would
be completely excluded. For $\tan \beta =2$ this would probe
approximately half the parameter space up to $M_{1/2}= 300$ GeV. 
Contrary to expectations the present bound of the $\tau \to \mu \gamma$ 
starts exploring the SUSY parameter already for low tan$\beta ~=2$. 
For larger tan $\beta~=40$ even the present bound rules out significant
regions of the parameter space. The main advantage of this decay mode is
that it does not depend on the value of $U_{e3}$ and therefore provides
an important constraint on the parameter space of the model for any
value of $U_{e3}$. 

It is important to emphasise that these radiative leptonic decays are a
much more powerful probe on the SUSY parameter space than the very similar 
$b \to s \gamma$ decay. This can be traced to the fact that slepton and 
chargino/neutralino masses are a factor $\sqrt{6}$ smaller that the
corresponding gluino and squark masses for the same GUT initial parameters
due to RGE effects. Therefore, these decays will become in the near future
the most stringent constraints on the SUSY parameter space and offer an
excellent opportunity for SUSY searches.

\section{Discussion of the results and final remarks} 
As we have mentioned, $SO(10)$ is an interesting and predictive scheme
for BR($l_j \rightarrow l_i,\gamma$) in so that one of the neutrino
Yukawa couplings has to be of the ${\mathcal O}(h_t)$. In our plots of
Figs 1 and 2 we exhibited results of the two `extreme' cases of
``small " (CKM-like) and ``large " (PMNS -like) lepton mixing
angles. We see that the constraints coming from BR($\mu \rightarrow
e,\gamma$) and BR($\tau \rightarrow \mu,\gamma$) are very
significant. In particular for the PMNS case, even the present bounds
on these two branching ratios (Eqs.~(\ref{megprelimit},
\ref{tmgprelimit})) are able to exclude large regions of the SUSY
$SO(10)$ parameter space which are still allowed by all the present
SUSY accelerator tests.

Needless to say, improving the experimental sensitivity of those two
processes, would place lepton radiative decays in the forefront of the
indirect searches of SUSY signals. Indeed, reaching ${\mathcal
O}(10^{-14})$ for BR($ \mu \rightarrow e, \gamma$) would probe larger
regions of the SUSY $SO(10)$ parameter space than that probed by
$b\rightarrow s, \gamma$ even in the less optimistic case of the small
CKM lepton angles. Obviously, this statement becomes stronger when one
moves to the PMNS case unless $U_{e3}$ is really very small, say
$U_{e3}~ \ler~ 10^{-3}$. If this latter circumstance occurs, then
$\tau \rightarrow \mu,\gamma$ becomes more powerful SUSY probe than
$\mu \rightarrow e,\gamma$ and again, reaching sensitivity for
BR($\tau \rightarrow \mu,\gamma$) of ${\mathcal O}(10^{-8})$ we could
do better than $b\rightarrow s, \gamma$ in constraining the SUSY
$SO(10)$ parameter space.

Our analysis, referring in all generality to the framework of a
relevant class of unified SUSY models, once again emphasises the
importance and need for a strenuous effort in pursuing the challenging
experimental road of lepton radiative decays.

\bigskip
\centerline{\bf Acknowledgements}
\smallskip
We acknowledge support from the RTN European project ``Physics Across
the Present Energy Frontier'' HPRN-CT-2000-0148.  O.V. acknowledges
partial support from the Spanish CICYT AEN99-0692 and DGEUI of the
Generalitat Valenciana under the Grant GV01-94.

\begin{appendix}
\begin{center}
{\textbf Appendix I}
\end{center}

In this appendix we discuss in detail the proof that the large hierarchy
in $M^u$ necessarily requires that the symmetric Yukawa couplings include
a large entry of ${\cal{O}}(m_t)$. Let us decompose a generic mass matrix 
in symmetric and anti-symmetric parts as follows: 
\be
M = M_S + M_A, 
\ee
where 
\be 
M_S = \left( \begin{array}{ccc}
a_{11} & a_{12}& a_{13} \\
a_{12} & a_{22} & a_{23} \\
a_{13} & a_{23} & a_{33} \end{array} \right) 
\ee 
and 
\be 
M_A = \left( \begin{array}{ccc}
0 & b_{12}& b_{13} \\
-b_{12} & 0 & b_{23} \\
-b_{13} & -b_{23} & 0 \end{array} \right) 
\ee 
The eigenvalues of these matrix can be found solving the cubic equation:
\be 
\lambda^3 + a \lambda^2 + b\lambda + c = 0 
\ee 
with the coefficients,
\bea
a &=& \mbox{Tr}[M^u M^{u~T}] \nn \\
\label{a1}
&=& a_{11}^2 + a_{22}^2 + a_{33}^2 + 2~( a_{12}^2 + a_{13}^2 + a_{23}^2 
+ b_{12}^2 + b_{23}^2 + b_{13}^2),
\eea
\bea
b &=& {\cal{M}}_{11}[M^u M^{u~T}] +
\label{b1}
{\mathcal{M}}_{22}[M^u M^{u~T}] + {\mathcal{M}}_{33}[M^u M^{u~T}] ~~~
\eea
where ${\mathcal{M}}_{ij}[M^u M^{u~T}]$ is the minor obtained by omitting
the $i$ row and $j$ column from the determinant. These are given as
\bea
{\mathcal M}_{11}&=& \left( a_{22}^2 + (b_{12} - a_{12})^2  + 
(a_{23} + b_{23})^2 \right) \left( (b_{13} - a_{13})^2 + 
a_{33}^2 + (b_{23} - a_{23})^2 \right) \nn \\
&-&  \left(a_{22}~ (a_{23} - b_{23}) + 
(a_{23} + b_{23})~ a_{33} + (b_{12} - a_{12}) (b_{13} - a_{13}) \right)^2 
\eea 
\bea
{\mathcal M}_{22}&=& \left( a_{11}^2 + (b_{12} + a_{12})^2  + 
(a_{13} + b_{13})^2 \right) \left( (b_{13} - a_{13})^2 + 
a_{33}^2 + (b_{23} - a_{23})^2 \right) \nn \\
&-&  \left(a_{11}~ (a_{13} - b_{13}) + 
(a_{13} + b_{13})~ a_{33} + (b_{12} + a_{12}) (a_{23} - b_{23}) \right)^2 
\eea 
\bea
{\mathcal M}_{33}&=& \left( a_{11}^2 + (b_{12} + a_{12})^2  + 
(a_{13} + b_{13})^2 ~\right) \left( (b_{12} - a_{12})^2 + 
a_{22}^2 + (b_{23} + a_{23})^2 \right) \nn \\
&-&  \left(a_{11}~ (a_{12} - b_{12}) + 
(a_{12} + b_{12})~ a_{12} + (a_{13} + b_{13}) (a_{23} + b_{23}) \right)^2 
\eea 
The parameter $c$ is given as follows: 
\bea
c &=& \mbox{Det}[M^u M^{u~T}] \nn \\
\label{c1}
&=&  \left( b_{23}^2~ a_{11} - 2~ b_{13}~ b_{23}~ a_{12} + 2~ b_{12}~ b_{23} 
~a_{13} + b_{13}^2 ~a_{22} - a_{13}^2 ~a_{22} - 
2~ b_{12} ~b_{13} ~a_{23} \right. \nn \\ 
&+& \left.  2 ~a_{12} ~a_{13} ~a_{23} - a_{11} ~a_{23}^2 + 
b_{12}^2 ~a_{33} - a_{12}^2 ~a_{33} + a_{11} ~a_{22} ~a_{33} \right)^2
\eea 
To generate a hierarchical spectrum, as required by the up-quark sector, 
$a,~b$ and $c$ have to satisfy the following conditions:
\bea
\label{trace}
a&=& m_u^2 ~+ m_c^2 +~ m_t^2 \\
\label{subdet}
b &=& m_u^2~ m_t^2 + m_c^2~ m_t^2 + m_u^2~ m_c^2 \\ 
\label{det}
c&=& m_t^2 m_c^2 m_u^2 
\eea 

We want to prove that at least one of the elements in the symmetric
matric must be of order $m_t$. To do this we show that the case where
$m_t$ has its origins solely from the anti-symmetric part
\textit{i.e,} the elements $b_{ij}$ is not consistent with the
observed spectrum. In such a case at least one of the $b_{ij}$ is as
large as $m_t$.  However from Eqs.~(\ref{a1}-\ref{c1}) we see that
such an assumption would be in conflict with the condition in
Eq.~(\ref{subdet}) as it leads to $b \sim m_t^4$ through one of the
${\mathcal M}_{ij}$. For ex: if $b_{23}^2 \sim m_t^2\;;\; {\mathcal
M}_{11} \sim m_t^4$. This is in fact true for any of the
$b_{ij}$. Therefore a large $b_{ij}$ would lead to a degenerate
spectrum rather than a hierarchical spectrum required. Thus $m_t$
cannot have its origins \textit{solely} from the anti-symmetric
part. The only allowed possibilities would then be, either $m_t$ has
its origins solely in the symmetric part 
or through elements of both $M_S$ and $M_A$,
when some elements in both matrices are of comparable magnitude $\sim~
{\mathcal O}(m_t)$.  In either case, it is clear that $M_S$ would
contain at-least one element as large as $m_t$.

\end{appendix}

\newpage

\begin{figure}[ht]
\includegraphics[scale=0.75]{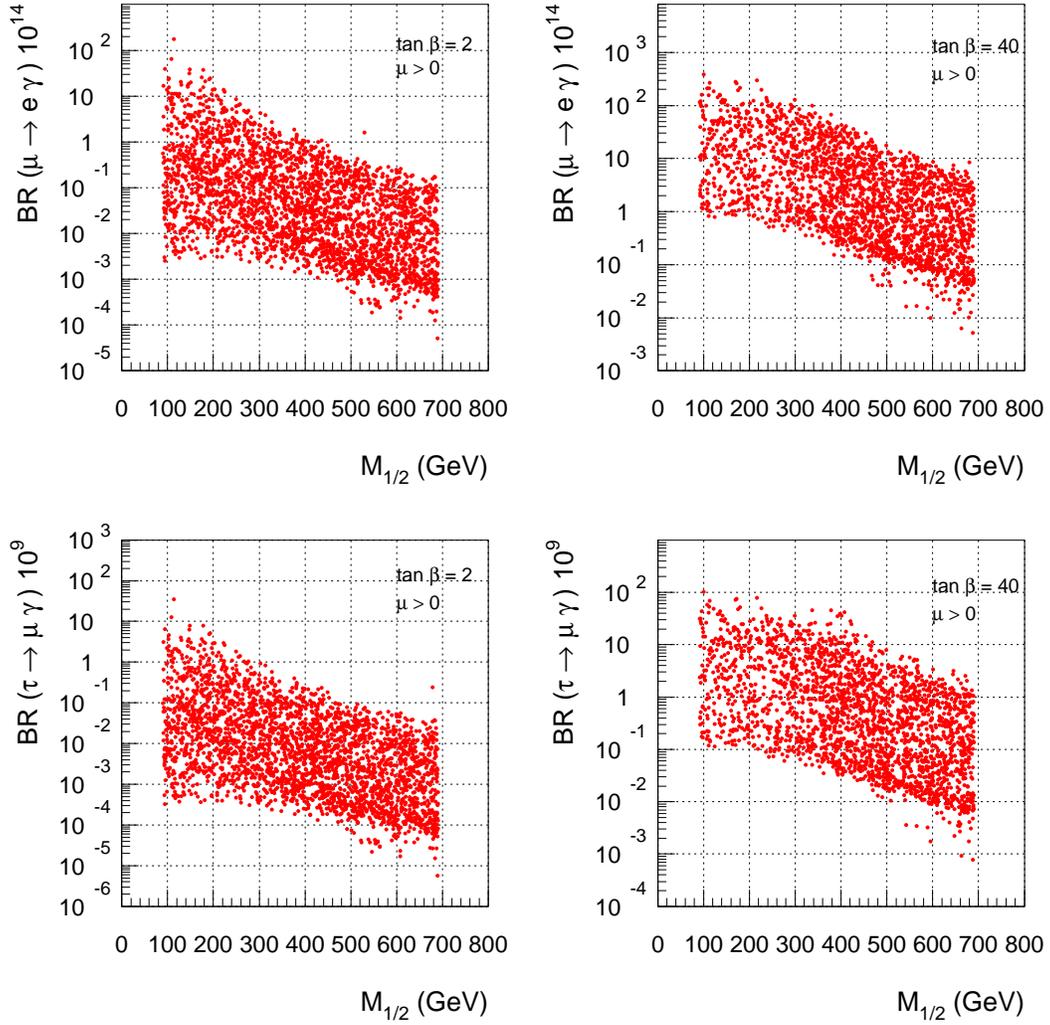}
\caption{The scatter plots of branching ratios of rare leptonic radiative
decays vs. $M_{1/2}$ are shown for the (minimal) CKM case for two specific 
values of tan $\beta$. Results do not alter significantly with the change 
of sign($\mu$).}
\end{figure}

\newpage

\begin{figure}[ht]
\includegraphics[scale=0.75]{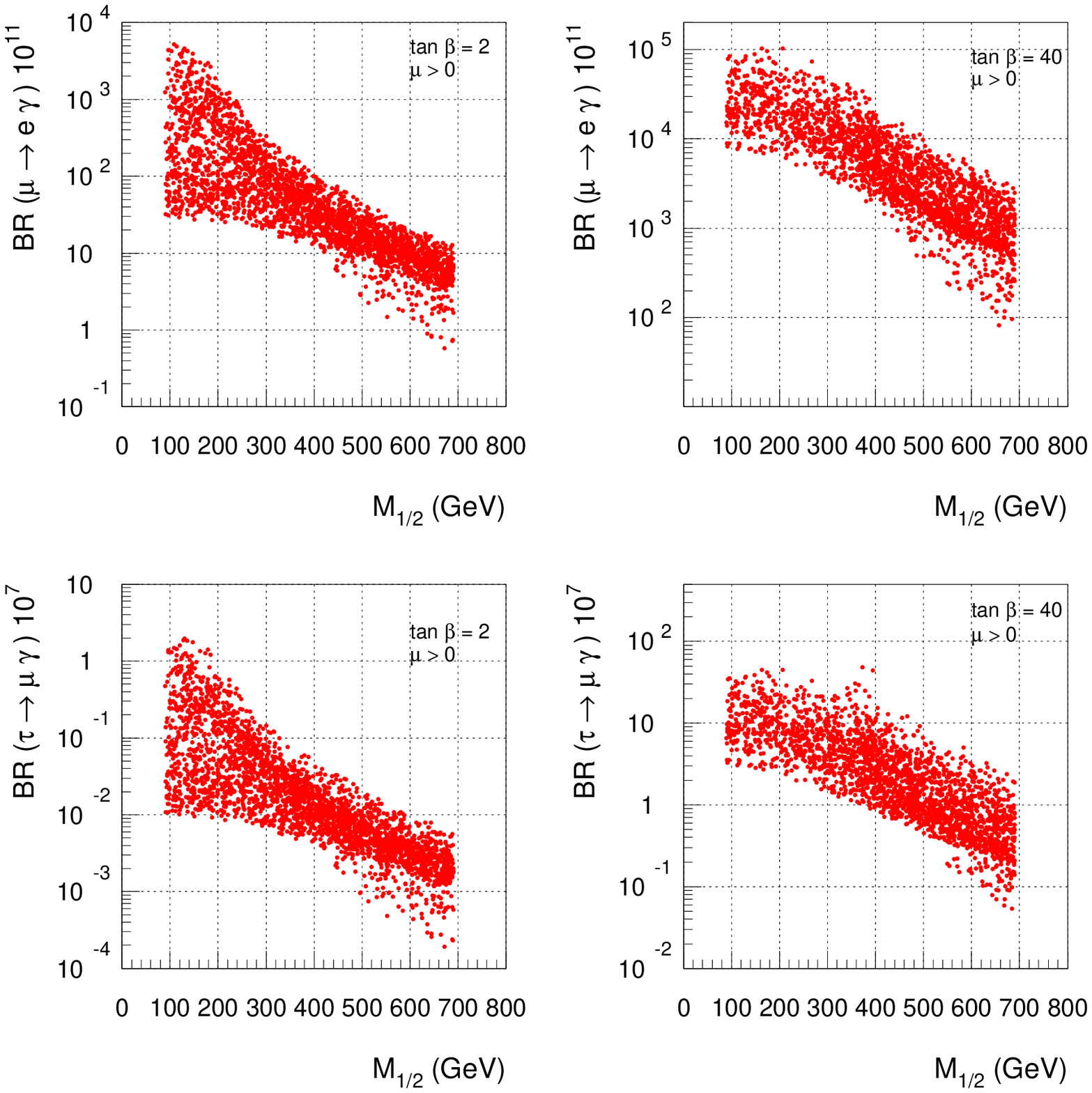}
\caption{The scatter plots of branching ratios of rare leptonic radiative
decays vs. $M_{1/2}$ are shown for the (maximal) PMNS case for two specific
values of tan $\beta$. Results do not alter significantly with the change 
of sign($\mu$).}
\end{figure}

\end{document}